\documentclass[11pt]{article}
\usepackage[margin=0.9in]{geometry}

\usepackage{slashed}
\usepackage{caption,subcaption}
\usepackage{amssymb,graphicx,xcolor,mathtools}
\usepackage{epstopdf}
\numberwithin{equation}{section}

\def\be {\begin{equation}}
\def\ee {\end{equation}}
\def\bea {\begin{eqnarray}}
\def\eea {\end{eqnarray}}
\def\bc {\begin{center}}
\def\ec {\end{center}}
\def\nn {\nonumber}

\def \tr{\operatorname{\text{tr}}}

\begin{document}
\title{Rho meson decay in presence of magnetic field}
\author{Aritra Bandyopadhyay\footnote{aritra.bandyopadhyay@saha.ac.in} and S. Mallik\footnote{mallik@theory.saha.ernet.in}}

\maketitle 
\bc
{\normalsize \it Theory Division,}\\
{\normalsize \it Saha Institute of Nuclear Physics, HBNI,} \\
{\normalsize \it 1/AF Bidhannagar, Kolkata 700064, India}
\ec
\begin{abstract}
We find a general expression for the one-loop self-energy function of neutral 
$\rho$-meson due to $\pi^+\pi^-$ intermediate state in a background magnetic field, 
valid for arbitrary magnitudes of the field.
The pion propagator used in this expression is given by Schwinger, which
depends on a proper-time parameter. Restricting to weak fields, we calculate the 
decay rate $\Gamma(\rho^0 \rightarrow \pi^+ +\pi^-)$, which changes negligibly 
from the vacuum value.
\end{abstract}

\section{Introduction}
\label{sec_intro}

%

Though some steller objects (like neutron star) were long known to possess magnetic fields~\cite{Woltjer,Wheaton,Fushiki}, 
the realization, that such fields are created in noncentral collisions of heavy ions~\cite{Kharzeev1,Skokov1,Skokov2,Skokov}, 
has initiated looking for effects of this background magnetic field on various observables~\cite{Fukushima1,Gatto,Ayala,Tuchin}. Thus its effect on dilepton production~\cite{Bandyopadhyay1,Bandyopadhyay2,Sadhoogi1} and on resonances created in the hadron phase~\cite{Chernodub1,Liu,Kawaguchi,Zhang,Ghosh} are investigated in detail. 
A more involved effect of this background field, called the chiral magnetic effect, demonstrates the 
topological nature of the QCD vacuum~\cite{Kharzeev1,Fukushima,Kharzeev2}. Apart from these effects in heavy ion collisions, the magnetic field can enhance the 
symmetry breaking of a theory, e.g. it increases the magnitude of the quark condensate, which breaks the flavor symmetry of QCD~\cite{Miransky,Mueller,Gusynin}.

Here we investigate the effect of an external magnetic field in the decay of $\rho$-meson in the dominant channel $\rho^0 \rightarrow \pi^+ + \pi^-$~\cite{Chernodub1}, which may affect the estimate of pion production in noncentral heavy ion collisions. This decay rate may be obtained from the imaginary part of the self energy graph of $\rho$-meson with two pion intermediate state (Fig \ref{rho_self_energy}). The effect of the external field can be included in the decay process by taking the modified pion propagation in this field.

Such a modified (scalar and spinor) propagator in coordinate space has been derived long ago by Schwinger~\cite{schwinger1}, to all orders in the external electromagnetic field, as an integral over proper time. Working in quantum electrodynamics, he used it to find corrections to 
Maxwell Lagrangian.
 But for the electron self energy function, he wrote the usual form, namely an integral over intermediate momentum $k_\mu$ 
 with the electron propagator that depends on the kinematical momentum $\Pi_\mu$, containing the electromagnetic potential $A_\mu$. 
 Then the shift of the origin in $k$ space, necessary in carrying out the $k$ integration, cannot be made, owing to the noncommutativity 
 of the components of $\Pi_\mu$. He circumvented this difficulty by an ingenius $\xi$-device\footnote{Ref~\cite{schwinger}, vol II, p. 224; vol III, p. 145} and evaluated the self energy function analytically for weak and strong magnetic fields. 
 
 In this work we write the pion self energy in coordinate space with pion propagators as given by Schwinger. There is no difficulty here as it contains no operators. The resulting expression is Fourier transformed to go over to momentum space. Having obtained the $\rho^0$ self energy for a general external field, we find its decay rate in a weak magnetic field. 
 
 In Section \ref{sec_pion} we outline Schwinger's derivation of scalar propagator in coordinate space. In Section \ref{sec_rho_se} we find the $\rho$ meson self energy, first in a general background field and then specialize it to magnetic field. In Section \ref{sec_rho_decay} we calculate the decay rate to order quadratic in the magnetic field. Finally a general discussion of our method is given in the last Section \ref{sec_conclu}. An Appendix evaluates the relevant integrals.

\section{Scalar Propagator}
\label{sec_pion}
The Lagrangian for charged pions of mass $m$ interacting with an external electromagnetic
field $A_\mu (x)$ is
\bea
\mathcal{L}&=& \left[(\partial_\mu+ieA_\mu)\phi\right]^\dagger(\partial^\mu+ieA^\mu)\phi-m^2\phi^\dagger\phi.
\label{lagrangian}
\eea
giving the equation of motion for the pion field as
\bea
\left[(\partial_\mu+ieA_\mu)(\partial^\mu+ieA^\mu)+m^2\right]\phi(x) = 0.
\label{eom1}
\eea
The pion propagator is defined as
\bea
G(x,x^\prime) = i\langle 0\vert T\phi(x)\phi^\dagger(x^\prime)\vert 0\rangle,
\eea
where $T$ represents the time ordering and $\vert 0\rangle$ is the vacuum
for the quantum fields. The propagator satisfies
\bea
\left[(\partial_\mu^x+ieA_\mu)(\partial^\mu_x+ieA^\mu)+m^2\right]G(x,x^\prime) = \delta^4(x-x^\prime).
\label{greens_fn}
\eea

We now review the steps arising in Schwinger's derivation of the exact
propagator. If we introduce states labeled by space-time coordinates, $G(x,x^\prime)$ may be written as the matrix element of an operator $G$
\bea
G(x,x^\prime) = \langle x^\prime\vert G\vert x\rangle,
\eea
when we can express Eq.(\ref{greens_fn}) as an operator equation
\bea
(-\Pi^2+m^2)G=1,
\label{operator_eq1}
\eea
where $\Pi_\mu = p_\mu - eA_\mu,~ p_\mu = i\partial_\mu$ and they satisfy the commutation relations,
\bea
\left[x^\mu,\Pi^\nu\right] = -ig^{\mu\nu}, ~~\left[\Pi^\mu,\Pi^\nu\right] = -ieF^{\mu\nu}. ~~~~(F^{\mu\nu} = \partial^\mu A^\nu-\partial^\nu A^\mu)
\eea
The operator equation (\ref{operator_eq1}) has the formal solution
\bea
G=\frac{1}{-\Pi^2+m^2}=i\int\limits_0^\infty ds~U(s),
\label{propagator_exponentiation}
\eea
where\footnote{Here $m^2$ is understood to be $m^2-i\epsilon$, the infinitesimal providing the convergence to $s$ integral and the boundary conditions for the time ordered propagator.}
\bea
U(s)=e^{-iHs},~~ H=-\Pi^2+m^2.
\eea
This notation emphasizes that $U(s)$ may be regarded as the operator describing the dynamics of a particle governed by the Hamiltonian $`H$' in the proper time parameter $`s$'. The spacetme coordinate $x^\mu = (t,\vec{x})$ of the particle depends on this parameter. \\

In the Heisenberg representation the operator $x_\mu$ and $\Pi_\mu$ have the $`$time' dependence
\bea
x_\mu(s) = U^\dagger(s)x_\mu U(s), ~~ \Pi_\mu(s) = U^\dagger(s)\Pi_\mu U(s),
\label{heisen_rep}
\eea
and the base ket and bra evolve as
\bea
\vert x^\prime ; s\rangle = U^\dagger(s)\vert x^\prime\rangle, ~~ \langle x^\prime ; s\vert = \langle x\vert U(s).
\eea
Then the construction of $G(x^\prime,x)$ reduces to the evaluation of
\bea
\langle x^{\prime\prime}\vert U(s)\vert x^\prime\rangle = \langle x^{\prime\prime} ; s\vert x^\prime ; 0\rangle,
\eea
which is the transformation function from a state in which the position operator $x_\mu(s=0)$ has value $x_\mu^\prime$, to a state in which $x_\mu(s)$ has the value $x_\mu^{\prime\prime}$. The equations of motion for the operators following from Eq.(\ref{heisen_rep}) are
\bea
\frac{dx_\mu}{ds} = -i\left[x_\mu,H\right],~~~~~\frac{d\Pi_\mu}{ds} = -i\left[\Pi_\mu,H\right].
\label{eom2}
\eea
The transformation function itself can be found by solving the differential equation satisfied by it,
\bea
i\frac{d}{ds} \langle x^{\prime\prime} ; s\vert x^\prime ; 0\rangle = \langle x^{\prime\prime} ; s\vert H(x(s),\Pi(s))\vert x^\prime ; 0\rangle,
\label{diff_eq1}
\eea
and
\bea
\left(i\partial_\mu^{x^{\prime\prime}}-eA_\mu(x^{\prime\prime})\right)\langle x^{\prime\prime} ; s\vert x^\prime ; 0\rangle = \langle x^{\prime\prime} ; s\vert \Pi_\mu(s)\vert x^\prime ; 0\rangle,
\label{diff_eq2}
\eea
along with a similar one for $\Pi_\mu(0)$, with boundary conditions
\bea
\langle x^{\prime\prime} ; s\vert x^\prime ; 0\rangle\Big\vert_{s\rightarrow 0} = \delta^4(x^{\prime\prime}-x^\prime), ~~ \lim_{s\rightarrow \infty}\langle x^{\prime\prime} ; s\vert x^\prime ; 0\rangle = 0.
\label{bdy_cond}
\eea

We now specialize to constant field strength, for which the eqs.(\ref{eom2}) can be solved exactly. Then Eqs.(\ref{eom2}) become, which in matrix notation reads
\bea
\frac{dx}{ds} = 2\Pi, ~~~~ \frac{d\Pi}{ds} = 2eF\Pi.
\label{eom_matrixform}
\eea
The second equation of (\ref{eom_matrixform}) can be immediately solved to give
\bea
\Pi(s) = e^{2eFs}\Pi(0).
\label{sol1}
\eea
With this solution the first equation of (\ref{eom_matrixform}) yields the solution
\bea
x(s)-x(0) = \frac{e^{2eFs}-1}{eF}\Pi(0).
\label{sol2}
\eea
Using (\ref{sol1}) and (\ref{sol2}) and the antisymmetry of the field tensor $(F^{\mu\nu}=-F^{\nu\mu})$ we get 
\bea
\Pi^2(s)=(x(s)-x(0))K(x(s)-x(0)), ~~ K=\frac{e^2F^2}{4}\left[\sinh(eFs)\right]^{-2}.
\eea
To evaluate the matrix element on the right of Eq.(\ref{diff_eq1}), we need to order the operator $x(s)$ to the left of $x(0)$, which would require the commutator
\bea
\left[x_\mu(s),x_\nu(0)\right] = i\left(\frac{e^{2eFs}-1}{eF}\right)_{\mu\nu}.
\eea
Then we get
\bea
\langle x^{\prime\prime};s\vert H(s)\vert x^{\prime};0\rangle = f(x^{\prime\prime},x^{\prime};s)\langle x^{\prime\prime};s\vert x^{\prime};0\rangle,
\eea
where
\bea
f=-(x^{\prime\prime}-x^{\prime})K(x^{\prime\prime}-x^{\prime})-\frac{i}{2}\tr\left[eF\coth(eFs)\right]+m^2,
\eea
with $\tr$ indicating the trace over $4\times 4$ matrices.
Eq.(\ref{diff_eq1}) can now be solved as
\bea
\langle x^{\prime\prime};s\vert x^{\prime};0\rangle &=& C(x^{\prime\prime},x^{\prime})\exp\left[-i\int ds^\prime f(x^{\prime\prime},x^{\prime};s)\right]\nn\\
&=& C(x^{\prime\prime},x^{\prime})\frac{1}{s^2}e^{-L(s)}\exp\left[-\frac{i}{4}(x^{\prime\prime}-x^{\prime})R(s)(x^{\prime\prime}-x^{\prime})-i(m^2-i\epsilon)s\right],
\eea
where\footnote{In writing $L(s)$ we follow Schwinger in choosing the integration constant to make $C$ independent of external field and also in extracting the coordinate independent, singular $s$ behavior.},
\bea
L(s)=\frac{1}{2}\tr\ln\left[(eFs)^{-1}\sinh(eFs)\right], ~~R(s) = eF\coth(eFs).
\eea
The $s$-independent function $C$ can be found by solving Eq.(\ref{diff_eq2})
\bea
C(x^{\prime\prime},x^{\prime})=C(x^{\prime\prime})\exp\left[-ie\int\limits_{x^{\prime}}^{x^{\prime\prime}}d\xi\left[A(\xi) +\frac{1}{2}F(\xi -x^\prime)\right]\right].
\eea
As $A(\xi) +\frac{1}{2}F(\xi -x^\prime)$ has vanishing curl, it can be written as
\bea
C(x^{\prime\prime},x^{\prime})=C~\Phi(x'',x')~;~~\Phi(x'',x') = \exp\left[-ie\int\limits_{x^{\prime}}^{x^{\prime\prime}}d\xi A(\xi)\right],
\eea
where the integration in the phase factor $\Phi$ runs on a straight line between $x'$ and $x''$ and the $F$ term vanishes. The constant $C$ is given by the first boundary condition of (\ref{bdy_cond}) as
\bea
1=\frac{C}{s^2}\int d^4x \exp\left(-\frac{i}{4s}x^2\right).
\label{contour_deformation}
\eea
This integral is evaluated in the Appendix, to give
$C=-i/(4\pi)^2$.\\
We finally get the transformation function as
\bea
\langle x^{\prime\prime};s\vert x^{\prime};0\rangle &=& -\frac{i}{(4\pi)^2s^2}~\Phi(x'',x')~e^{-L(s)}\times\nn\\
&&\exp\left[-\frac{i}{4}(x^{\prime\prime}-x^{\prime})R(s)(x^{\prime\prime}-x^{\prime})-i(m^2-i\epsilon)s\right],
\label{transformation_function}
\eea
which in turn gives the propagator
\bea
G(x^{\prime\prime},x^\prime) = i\int\limits_0^\infty ds~ \langle x^{\prime\prime};s\vert x^{\prime};0\rangle.
\label{pion_propagator}
\eea
In our work below we shall encounter the product of propagators, $G(x',x'')G(x'',x')$ with derivatives acting on them. Without the derivatives, the phase factor in $G$ would cancel out mutually. In presence of derivatives we can still get rid of the phase factors, if we make a gauge choice in the potential, replacing $A_\mu$ with 
\be
A'_\mu(x) = A_\mu(x)-\partial_\mu\lambda(x), ~~\lambda(x) = \int\limits_{x'}^x d\xi^\mu~A_\mu(\xi),
\ee
when Eq.(\ref{greens_fn}) is satisfied by $G$ without the phase factor $\Phi$~\footnote{Ref~\cite{schwinger}, vol I, p. 271;}. In the following we choose this gauge to write $G(x'',x')$ without this phase.

\section{$\rho$ self-energy in external field}
\label{sec_rho_se}

We now express the self energy graph of Fig. \ref{rho_self_energy} in terms of pion propagators in external field given by Eqs.(\ref{transformation_function}) and (\ref{pion_propagator}). We first keep the external (constant) field a general one, specializing later to the interesting case of pure magnetic field.

\begin{center}
 \begin{figure}
 \begin{center}
  \includegraphics[scale=0.6]{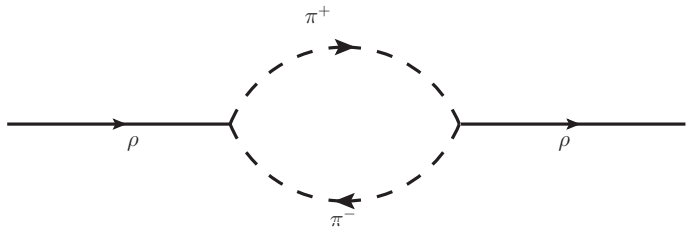}
  \caption{ Self energy graph of $\rho$-meson from two-pion intermediate state in presence of magnetic field of arbitrary strength.}
  \label{rho_se}
  \end{center}
 \end{figure}
\end{center}

\subsection{General field}

We take a phenomenological Lagrangian for$\rho\pi\pi$ interaction, 
\bea
\mathcal{L}_{int} = ig\rho^\mu(x)\phi^\dagger(x)\overleftrightarrow{\partial_\mu}\phi(x),
\label{int_lagrang}
\eea
The coupling $g$ can be found from the experimental decay width 
$\Gamma(\rho^0\rightarrow\pi^+\pi^-)=149$ MeV in vacuum to give $g=6.0$.
We now work out the complete $\rho$-propagator
\bea
D^\prime_{\lambda\sigma}(z,z^\prime) = i\langle0\vert T\rho_\lambda(z)\rho_\sigma(z^\prime)\exp\left(i\int dx~\mathcal{L}_{int}(x)\right)\vert 0\rangle,\nn
\eea
to order $g^2$ in the interaction representation. We take the $\rho$ field as free\footnote{The $\rho$ meson would acquire an external field dependent mass~\cite{Chernodub1}. But we do not include it here, as we are interested in the imaginary part of the self energy.}, but the pion field lives in the background electromagnetic field. Contracting the fields for graph of Fig. \ref{rho_se}, we get 
\bea
D^\prime_{\lambda\sigma}(z-z^\prime) = D_{\lambda\sigma}(z-z^\prime) + \int d^4x^{\prime}d^4x^{\prime\prime}D_{\lambda\mu}(z-x^\prime)\Sigma^{\mu\nu}(x^{\prime}-x^{\prime\prime})D_{\nu\sigma}(x^{\prime\prime} -z),
\label{DSeqn}
\eea
where $\Sigma$ is the self-energy tensor involving the two pion propagators. The propagators are distinguished only by their proper times $s_1$ and $s_2$, over which they are integrated. Carrying out the derivatives contained in $\mathcal{L}_{int}$ on these propagators, we get~\cite{schwinger2,yildiz} 
\bea
\Sigma^{\mu\nu}(x) &=& -i\left(\frac{g}{32\pi^2}\right)^2\int \frac{ds_1ds_2}{s_1^2s_2^2}\exp\left[-L_1-L_2\right] ~\Lambda^{\mu\nu}(x)
\label{self_energy_coord}
\eea
with
\bea
\Lambda^{\mu\nu}(x) &=& \left[x_\alpha(R_1-R_2)^{\alpha\mu}(R_1-R_2)^{\nu\beta}x_\beta +2i(R_1+R_2)^{\mu\nu}\right]\times\nn\\
&& \exp\left[-\frac{i}{4} x_\alpha(R_1+R_2)^{\alpha\beta}x_\beta-im^2(s_1+s_2)\right].
\label{rho_self_energy}
\eea
Here $L_1=L(s_1), L_2=L(s_2)$ and similarly for $R_1, R_2$. 

Having obtained equation (\ref{DSeqn}) in configuration space, we go over to momentum space by taking Fourier transforms. Letting $K_{\mu\nu}(x)$ to denote any of the $D_{\mu\nu}(x), D^\prime_{\mu\nu}(x)$ and $\Sigma_{\mu\nu}(x)$, their Fourier transforms are defined as
\bea
K_{\mu\nu}(x) = \int\frac{d^4q}{(2\pi)^4}~e^{-i~q\cdot x}K_{\mu\nu}(q)\nn
\eea
and Eq. (\ref{DSeqn}) becomes 
\bea
D^\prime_{\lambda\sigma}(q) = D_{\lambda\sigma}(q) + D_{\lambda\mu}(q)\Sigma^{\mu\nu}(q)D_{\nu\sigma}(q).
\label{DSeqn_momspace}
\eea
The vacuum and the complete propagator of $\rho$ meson are given by
\bea
D_{\lambda\sigma}(q) &=& \left(-g_{\lambda\sigma}+\frac{q_\lambda q_\sigma}{m^2}\right)D(q^2),~~D(q^2) = \frac{-1}{q^2-m_\rho^2+i\epsilon},\\
D^\prime_{\lambda\sigma}(q) &=& \left(-g_{\lambda\sigma}+\frac{q_\lambda q_\sigma}{m^2}\right)D^\prime(q^2),
\eea
where $D'(q^2)$ is the function we want to find.

A simplification results on noting that the $\rho$ field is coupled to a conserved pion current in the interaction lagrangian given by Eq. (\ref{int_lagrang}). As a result, contracting $q_\mu$ and $q_\nu$ of the $\rho$ propagators with $\Sigma^{\mu\nu}$ in the second term of Eq.(\ref{DSeqn_momspace}) yield zero. We are then left with the metric tensor in $\rho$ propagator. Contracting further the indices $\sigma$ and $\lambda$ we get
\bea
-3D^\prime(q^2) = -3D(q^2) + D(q^2)\Sigma(q)D(q^2).
\label{reduced_rho_prop}
\eea
Here $\Sigma(q)$ is the Fourier transform of $\Sigma^{\mu\nu}(x)$ after contracting the indices,
\bea
\Sigma(q) = -i\left(\frac{g}{32\pi^2}\right)^2 \int \frac{ds_1ds_2}{s_1^2s_2^2}\exp\left[-L_1-L_2\right]~\Lambda(q)
\label{sigma_msp}
\eea
with
\bea
\Lambda(q) &=& \exp\left[-im^2(s_1+s_2)\right] \int d^4x \left[x^\rho(R_1-R_2)^2_{\rho\sigma} x^\sigma +2ig_{\mu\nu}(R_1+R_2)^{\mu\nu}\right]\times\nn\\
&&\exp\left[iq\cdot x-\frac{i}{4}x^\alpha(R_1+R_2)_{\alpha\beta}x^\beta-im^2(s_1+s_2)\right].
\label{scaler_self_energy}
\eea
Clearly the expression for $\Sigma$ is divergent, to which we have to add renormalization counterterms. Beside cancelling the divergent pieces, we shall choose the finite pieces in the counterterms, such that the total self energy $\Sigma_{tot}$ satisfies 
\bea
\Sigma_{tot}(q^2=m_\rho^2) = 0;~~~ \frac{d\Sigma_{tot}(q)}{dq}\Bigg\vert_{q^2 = m_\rho^2}=0.\nn
\eea
Then $m_\rho$ will remain the physical $\rho$-meson mass and $g$, the renormalized $\rho\pi\pi$ coupling. We shall come back to this renormalization in Section \ref{sec_rho_decay} below.

Including the sum of all reducible graphs in Eq.(\ref{reduced_rho_prop}) we get the Dyson-Schwinger equation for the $\rho$ propagator as
\bea
D^\prime(q^2) = D(q^2) - \frac{1}{3} D(q^2) \Sigma(q) D^\prime(q^2),
\eea
giving
\bea
D^\prime (q^2) = \frac{-1}{q^2-m_\rho^2-\frac{1}{3}\Sigma(q)}.
\label{complete_prop}
\eea
In the neighbourhood of physical $\rho$-meson pole, it gives the decay width $\Gamma(\rho^0 \rightarrow \pi^+\pi^-)$ with background electromagnetic field as
\bea
\Gamma = -\frac{\text{Im}\Sigma}{3m_\rho}.
\label{decay_width}
\eea
Below we shall calculate $\text{Im}\Sigma$.

\subsection{Pure magnetic field}

We derived above the expressions for the pion propagator and the consequent $\rho$-meson self energy in a general background external field $F_{\mu\nu}$. We now specialize this field to magnetic field $B$ in the $z$ direction, i.e. $F^{21}=-F^{12}=B$ and all other components zero. We can diagonalize this $2\times 2$ matrix, in which only the coordinates $x^{\prime 1}$ and $x^{\prime 2}$ are involved. After diagonalization,  
the quantities involving matrices reduce to,
\bea
L(s) &=& \ln\left[(eBs)^{-1}\sin(eBs)\right],~~R=2eB\cot(eBs)+\frac{2}{s},\nn\\
(x^{\prime\prime}-x^{\prime})R(s)(x^{\prime\prime}-x^{\prime}) &=& -(x^{\prime\prime}-x^{\prime})_{\perp}^2eB\cot(eBs) + (x^{\prime\prime}-x^{\prime})_{\shortparallel}^2 \frac{1}{s},
\label{magfield_evaluations}
\eea
where $(x^{\prime}_{\perp})^2 = (x^{\prime~1})^2+(x^{\prime~2})^2$; $(x^{\prime}_{\shortparallel})^2 = (x^{\prime~0})^2-(x^{\prime~3})^2$. 
With these values, Eqs.(\ref{transformation_function}) and (\ref{pion_propagator}) give the pion propagator in the magnetic field as
\bea
G(x) &=& \frac{eB}{(4\pi)^2}~\int\limits_0^\infty~\frac{ds}{s~\sin(eBs)}~\exp\left[ix_{\perp}^2\frac{eB}{4}\cot(eBs)-\frac{i}{4s}x_{\shortparallel}^2-i(m^2-i\epsilon)s\right].
\label{transformation_function_magnetic}
\eea
Then the corresponding $\rho$ self energy in momentum space can be obtained from Eqs. (\ref{sigma_msp}) amd (\ref{scaler_self_energy}). It will involve two integrals over the proper time $s_1$ and $s_2$ of the propagating pions. We change these variables to $s$ and $u$  defined by
\bea
s_1=su~,~~ s_2=s(1-u)~,~~ ds_1~ds_2 = s~ds~du.
\label{change_of_variables}
\eea
Introducing for short
\bea
\alpha &=& (eB)^2 \left[\cot(eBsu)-\cot(eBs(1-u))\right]^2,\nn\\
\beta &=& \frac{1}{s^2}\left(\frac{1}{u}-\frac{1}{(1-u)}\right)^2,\nn\\
\gamma &=& \frac{eB}{4} \left[\cot(eBsu)+\cot(eBs(1-u))\right],\nn\\
\delta &=& \frac{1}{4su(1-u)}.
\label{short_forms}
\eea
we can now write the self energy as
\bea
\Sigma(q_\perp,q_\shortparallel) = -i\left(\frac{g}{32\pi^2}\right)^2 \int\limits_0^\infty \frac{ds}{s}\int\limits_0^1 \frac{du}{u(1-u)} ~\frac{(eB)^2 \Lambda(s,u)}{\sin(eBsu)\sin(eBs(1-u))},
\label{self_energy_su}
\eea
with 
\bea
\Lambda(s,u,q_\shortparallel,q_\perp) &=& e^{-im^2s}\int d^4 x e^{i q\cdot x}\left[-\alpha x_\perp^2 +\beta x_\shortparallel^2+16 i (\gamma +\delta)\right]\exp\left(i\gamma x_\perp^2 - i\delta x_\shortparallel^2\right),
\label{lambda_su}
\eea
which is evaluated in the Appendix yielding
\bea
\Lambda (s,u) &=& \frac{\pi^2}{\gamma\delta}\left[\frac{\alpha}{\gamma}\left(1-\frac{iq_\perp^2}{4\gamma}\right)+\frac{\beta}{\delta}\left(1+\frac{iq_\shortparallel^2}{4\delta}\right)-16(\gamma +\delta)\right]\nn\\ &&\times\exp\left[-i\frac{q_\perp^2}{4\gamma}-i\left\{m^2-q_\shortparallel^2u(1-u)-i\epsilon\right\}s\right].
\label{lambda_fap}
\eea
Note that all the parameters $\alpha,\beta,\gamma$ and $\delta$ are positive. 

The $i\epsilon$ prescription in Eq.(\ref{lambda_su}) makes the integration line in the $s$-plane run infinitesimally below the real axis, avoiding the singularities of the integrand. Then we can get rid of oscillations in it by deforming the integration line. As $u(1-u)\le 1/4$ for $0<u<1$, the quantity $\left[m^2-q_\shortparallel^2u(1-u)\right]$ appearing in Eq.(\ref{lambda_su}) is positive, if we take $q_\shortparallel^2 < 4m^2$, precluding physical pion pair creation. For such values of $q_\shortparallel^2$ we take a closed contour in the fourth quadrant, with vanishing contribution from the quarter circle\footnote{Ref~\cite{Itzykson}, p. 322;}. Changing the integration variable $s=-it$ on the imaginary axis, we get
\bea
\Sigma(q_\perp,q_\shortparallel) &=& \left(\frac{g}{32\pi}\right)^2 \int\limits_0^\infty \frac{dt}{t}\int\limits_0^1 \frac{du}{u(1-u)}~\frac{(eB)^2 \Lambda^\prime(t,u)}{\sinh(eBtu)\sinh(eBt(1-u))},
\label{self_energy_tu}
\eea
where $\Lambda^\prime$ is related to $\Lambda$ by the change of variable. To write $\Lambda^\prime$, we define a new set of variables from Eq.(\ref{short_forms}),
\bea
\alpha^\prime &=& (eB)^2 \left[\coth(eBtu)-\coth(eBt(1-u))\right]^2,\nn\\
\beta^\prime &=& \frac{1}{t^2}~\frac{(1-2u)^2}{u^2(1-u)^2},\nn\\
\gamma^\prime &=& \frac{eB}{4} \left[\coth(eBtu)+\coth(eBt(1-u))\right],\nn\\
\delta^\prime &=& \frac{1}{4tu(1-u)}.
\eea
In terms of these variables, we have
\bea
\Lambda^\prime &=& \frac{1}{\gamma^\prime\delta^\prime}\left[\frac{\alpha^\prime}{\gamma^\prime}\left(1-\frac{q_\perp^2}{4\gamma'}\right)+\frac{\beta^\prime}{\delta^\prime}\left(1 +\frac{q_\shortparallel^2}{4\delta^\prime}\right)-16(\gamma^\prime+\delta^\prime)\right]\nn\\
&&~~~~~~~~~\times~~~~\exp\left[-\frac{q_\perp^2}{4\gamma'}-\left\{m^2-q_\shortparallel^2u(1-u)\right\}t\right].
\label{lambda_prime}
\eea

Another form of $\Sigma$, which will be useful for the discussion below may be obtained by scaling $t$ with $eB$, that is, we set $\bar{t} = eBt$, when $\Sigma$ becomes
\bea
\Sigma = \int\limits_0^\infty\frac{d\bar{t}}{\bar{t}}\int\limits_0^1\frac{du}{u(1-u)}\frac{\bar{\Lambda}}{\sinh(\bar{t}u)\sinh(\bar{t}(1-u))},
\label{self_energy_sc}
\eea
where 
\bea
\bar{\Lambda} &=& \frac{1}{\bar{\gamma}\bar{\delta}}\left[\frac{\bar{\alpha}}{\bar{\gamma}}\left(eB-\frac{q_\perp^2}{4\bar{\gamma}}\right)+\frac{\bar{\beta}}{\bar{\delta}}\left(eB +\frac{q_\shortparallel^2}{4\bar{\delta}}\right)-16eB(\bar{\gamma}+\bar{\delta})\right]\nn\\
&&\times \exp\left[-\frac{q_\perp^2}{4\bar{\gamma}}-\left\{\frac{m^2}{eB}-\frac{q^2}{eB}u(1-u)\right\}t\right],
\label{lambda_bar}
\eea
with
\bea
\bar{\alpha} &=& \left[\cot(\bar{t}u)-\cot(\bar{t}(1-u))\right]^2,~~\bar{\beta} = \frac{1}{\bar{t}^2}\left(\frac{1}{u}-\frac{1}{(1-u)}\right)^2,\nn\\
\bar{\gamma} &=& \frac{1}{4} \left[\cot(\bar{t}u)+\cot(\bar{t}(1-u))\right],~~\bar{\delta} = \frac{1}{4\bar{t}u(1-u)}.
\label{short_forms_bar}
\eea

\section{$\rho$-meson decay}
\label{sec_rho_decay}

As already stated, the self energy function in Eq.(\ref{self_energy_tu}) is valid 
for momenta, for which the $\rho$-meson cannot decay into a pion pair. To calculate 
this decay rate we therefore need to continue Eq.(\ref{self_energy_tu}) beyond such 
momenta~\cite{Itzykson}. This process of analytic continuation is immediate, if we can evaluate the 
$t$ integral analytically. However the exact expression (\ref{self_energy_sc}) for $\Sigma$
containing various hyperbolic functions, makes it difficult to do so. The procedure here 
is to consider separately weak ($eB<m^2$) and strong ($eB>m^2$) fields\footnote{Ref~\cite{schwinger}, vol III, p. 161-163;}. 
As strong fields are considered extensively in the literature~\cite{Chernodub1,Liu,Kawaguchi,Zhang}, we take up 
the case of weak fields, which is realized, in particular, in the hadronic phase of noncentral heavy ion collisions. In this case the exponential in Eq.(\ref{lambda_bar}) 
shows that only correspondingly small values of $t$ 
can contribute. We can then expand the different functions in powers of $t$.
 To get the leading effect, we need to keep only the first 
two terms in their expansions. After some algebra, we get the self energy as
\bea
\Sigma(q^2) &=& \left(\frac{g}{2\pi}\right)^2\frac{1}{2} \int\limits_0^\infty dt\int\limits_{-1}^1 
dv (D+\frac{(eB)^2}{6}F)~\exp\left[-\left\{m^2-\frac{q^2}{4}(1-v^2)\right\}t\right],
\label{self_energy_weak}
\eea
where
\bea
D &=& \frac{v^2}{t}\frac{q^2}{4}-\frac{2}{t^2}, \nn\\
F &=& \frac{3}{2}(1-v^2)-\frac{v^2}{4}\left\{q^2-2(1-v^2)q_\perp^2\right\}t.
\label{DF}
\eea
Here a numerical factor of $4^4$ has been put into the coupling constant factor. Also for 
convenience, the variable $u$ is replaced by $\frac{1}{2}(1+v)$. 

As the terms in $D$ do not depend on the magnetic field and we want to calculate the change in $\rho$ meson decay width in this magnetic field, it is clear that our calculation will not involve $D$. However we want to show the nature of those terms. To this end, consider the first term in $D$, behaving as $t^{-1}$. Integrating partially w.r.t $v$, it gives
\bea
&&\left(\frac{g}{2\pi}\right)^2\frac{q^2}{4} \int\limits_0^1 dv~v^2 \int\limits_0^\infty \frac{dt}{t}~\exp\left[-\left\{m^2-\frac{q^2}{4}(1-v^2)\right\}t\right]\nn\\
&=& \left(\frac{g}{2\pi}\right)^2\frac{q^2}{12}\left[\int\limits_0^\infty \frac{dt}{t}e^{-m^2t}+\frac{q^2}{2}\int\limits_0^1 dv~v^4 \int\limits_0^\infty dt~\exp\left[-\left\{m^2-\frac{q^2}{4}(1-v^2)\right\}t\right]\right]
\eea
where the divergence at $t=0$ is isolated in the first term. The second term in $D$ behaving as $t^{-2}$ can also be put in a similar form after integrating twice partially w.r.t $v$. These are the local divergent terms, which we expected in Section \ref{sec_rho_se} for the general field case and are analogous to those appearing in loop integrals over intermediate momenta in conventional field theory. As a consistency check in our calculation, let us note that though individual terms in $\Lambda'$ given by Eq.(\ref{lambda_prime}) do contain $(eB)^2$ dependent divergent terms, they cancel out in the complete expression for $\Lambda'$, showing that divergences originate only from the vacuum piece of self energy, as expected.

Going back to the $eB$ dependent self energy given by the $F$ terms in Eq.(\ref{DF}), we rewrite it as
\bea
\Sigma_{eB}(q^2)=\left(\frac{g}{2\pi}\right)^2\frac{(eB)^2}{6}\overline{\Sigma}(q^2)
\label{se_eB}
\eea
where
\bea
\overline{\Sigma}(q^2) &=& \int\limits_0^1 dv\left[\frac{3}{2}(1-v^2)+\frac{v^2}{4}\left\{q^2-2(1-v^2)q_\perp^2\right\}\frac{\partial}{\partial m^2}\right]\nn\\
&\times& \int\limits_0^\infty dt ~\exp\left[-\left\{m^2-\frac{q^2}{4}(1-v^2)\right\}t\right].
\eea
We are now in a position to carry out the analytic continuation mentioned at the beginning of the section. If we hold the $q^2$ variable in the region $q^2<4m^2$, the $t$ integration is well defined and can be integrated trivially to give
\bea
\overline{\Sigma}(q^2)&=& \frac{3}{2}\int\limits_0^1 dv\frac{1-v^2}{m^2-q^2(1-v^2)}\nn\\ 
&+& \frac{\partial}{\partial m^2} \int\limits_0^1 dv\left[\frac{v^2}{4}\left\{q^2-2(1-v^2)q_\perp^2\right\}\right]\frac{1}{m^2-q^2(1-v^2)}.
\eea
It can now be continued for $q^2<4m^2$ in the $q^2$ plane with a cut along the real axis for $4m^2<q^2<\infty$. To display this cut structure explicitly, we write $\overline{\Sigma}(q^2)$ as a dispersion integral by changing the integration variable $v$ to $q'^2$ given by $v=\sqrt{1-4m^2/q'^2}$, getting
\bea
\overline{\Sigma}(q^2) &=& \int\limits_{4m^2}^\infty\frac{dq'^2}{q'^2}\frac{1}{q'^2-q^2}\left(1+\frac{12m^2}{q'^2}\right)\left(1-\frac{4m^2}{q'^2}\right)^{-1/2}\nn\\
&+& \frac{1}{2} \frac{\partial}{\partial m^2} \int\limits_{4m^2}^\infty\frac{dq'^2}{q'^2}\frac{1}{q'^2-q^2}\left(q^2-\frac{8m^2}{q'^2}q_\perp^2\right)\left(1-\frac{4m^2}{q'^2}\right)^{1/2}.
\eea
It's imaginary part is given by the discontinuity across the cut 
\bea
\text{Im} \overline{\Sigma}(q^2) &=& \frac{1}{2i}\left[\overline{\Sigma}(q^2+i\epsilon)-\overline{\Sigma}(q^2-i\epsilon)\right], ~~q^2>4m^2 \nn\\
&=& -\frac{\pi}{q^2}\left[\left(1-\frac{12m^2}{q^2}-\frac{8m^2}{q^2}\frac{q_\perp^2}{q^2}\right)\left(1-\frac{4m^2}{q^2}\right)^{-1/2}+\frac{4q_\perp^2}{q^2}\left(1-\frac{4m^2}{q^2}\right)^{1/2}\right].
\label{se_eB_im}
\eea
From Eqs.(\ref{decay_width}), (\ref{se_eB}) and (\ref{se_eB_im}), we get the corresponding change in width as
\bea
\Gamma_{eB} = \frac{g^2}{4\pi}\frac{(eB)^2}{18m_\rho^3}\left[1-10\frac{m^2}{m_\rho^2}+4\frac{q_\perp^2}{m_\rho^2}\left(1-\frac{4m^2}{m_\rho^2}\right)+\mathcal{O}\left(\frac{m^2}{m_\rho^2}\right)^2\right].
\label{Gma_eb}
\eea
Taking $m^2/m_\rho^2=1/30$, it becomes
\bea
\Gamma_{eB} = \frac{g^2}{4\pi}~\frac{(eB)^2}{27m_\rho^3}~\left(1+\frac{26}{5}\frac{q_\perp^2}{m_\rho^2}\right).
\eea
For $eB<m^2$ and $q_\perp^2<m_\rho^2$, it gives $\Gamma_{eB}<0.6$ MeV. The smallness of $\Gamma_{eB}$ may be explained by the fact that while the (small) pion mass is the scale entering in the self energy loop, it is evaluated at a (large) external momentum of $\rho$ meson mass. Also note that there is no pion mass in the denominator of Eq.(\ref{se_eB_im}). It is protected by chiral symmetry ($m\rightarrow 0$), according to which physical quantities must be finite in this limit. 

In passing, we note that the effect of temperature on the decay width of $\rho$-meson has been discussed extensively in the literature~\cite{Rapp}. Here we have investigated the effect of weak magnetic fields on the same quantity and found it to be negligible with respect to the thermal effects.

\section{Discussion}
\label{sec_conclu}

In earlier calculations of hadron properties in a magnetic field~\cite{Chernodub1,Liu,Kawaguchi,Zhang},
the majority of works consider strong fields, taking the contribution of the leading Landau level 
for the system. A result to note at this point is that for strong enough fields the main decay channel, 
such as $\rho^0 \rightarrow \pi^+ + \pi^-$ that we are considering, may become closed. It is due to the 
generation of an effective pion mass $\bar{m}^2 = m^2 +eB$, causing the phase space for the process 
to shrink as the magnetic field becomes stronger~\cite{Chernodub1}.

In the present work we investigate the decay by setting up a general framework, valid for both weak 
and strong magnetic fields. It is obtained by writing the $\pi\pi$ loop in the correction to the 
$\rho$ propagator in configuration space, with pion propagator as given by Schwinger~\cite{schwinger1}. 
When Fourier transformed, it gives the $\rho$ meson self energy (\ref{self_energy_tu}) as an 
integral over proper times, which is defined for momenta below the two-pion threshold. 

If we now restrict the general representation Eq.(\ref{self_energy_tu}) to weak fields ($eB<m$), 
the exponential factor in it (or quivalently Eq.(\ref{lambda_bar})) shows the leading contribution to arise from the neighbourhood of proper 
time $t=0$, when we can expand the hyperbolic functions in powers of $t$. Still remaining below the 
two-pion threshold, we can integrate the resulting terms to get a series in powers of $(eB)^2$. These 
terms can be simply continued beyond the threshold and the imaginary part of the self energy giving the 
decay width can be determined. In this work we retain only the $(eB)^2$ terms, though calculation of 
higher order terms is also straightforward. As we show at the end of Section \ref{sec_rho_decay}, the 
change in the decay width from the vacuum value turns out to be negligibly small.

So far we only discussed the effect of weak magnetic fields. But as already emphasized, strong field 
effects can also be obtained from the same general formula Eq.(\ref{self_energy_tu}). For $eB>m^2$,
 the exponential in this formula shows that large values of $t$ would also contribute. It is thus simple 
 to keep the leading term in different hyperbolic functions. Collecting the exponentials in 
 Eq.(\ref{self_energy_tu}), we get 
 \bea
 \exp\left[-\frac{q_\perp^2}{2eB} - \{m^2+eB-q_\shortparallel^2u(1-u)\}t\right]\nn 
 \eea
 giving the effective pion mass, as mentioned above.
 
 There are at least two other methods of calculating the decay rate. One is Schwinger's $\xi$-device mentioned 
 in Section \ref{sec_intro} and the other is the Ritus method of eigenfunction expansion~\cite{Ritus}. It will 
 be interesting to get comparable values from these methods.

\section{Acknowledgement}

The work of AB is supported by the Depatment of Atomic Energy (DAE), India.

\appendix

\section{Integrals}

\begin{center}
 \begin{figure}
 \begin{center}
 \begin{subfigure}[b]{0.45\textwidth}
        \centering
        \includegraphics[scale=0.5]{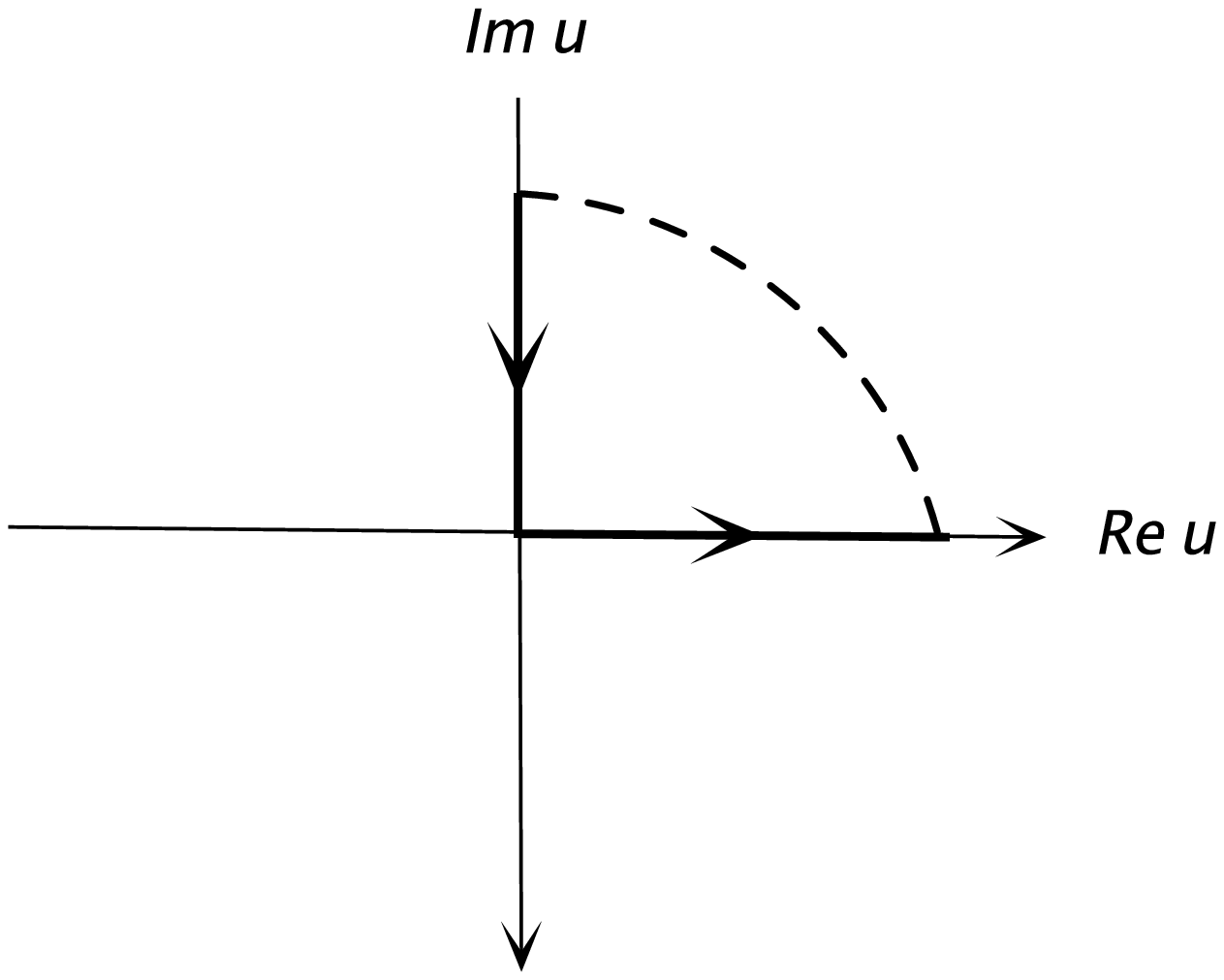}
        \caption{~~}
        \label{contour_mf1}
    \end{subfigure}
    \begin{subfigure}[b]{0.45\textwidth}
        \centering
        \includegraphics[scale=0.5]{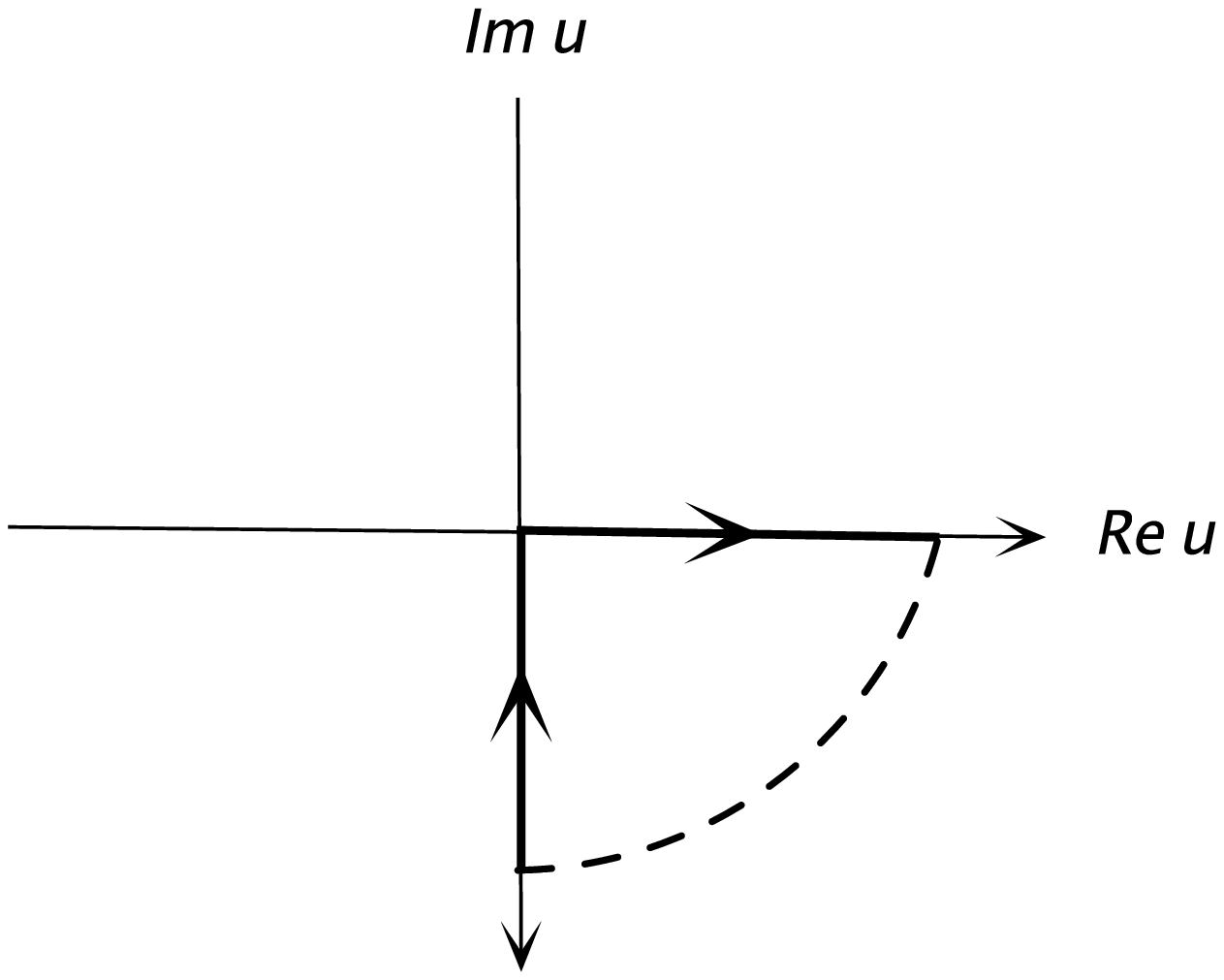}
        \caption{~~}
        \label{contour_mf2}
    \end{subfigure}
\caption{Closed contour in $u$ plane in both first and fourth quadrants.}
 \end{center}
 \end{figure}
\end{center}

Here we evaluate the integrals in Eq. (\ref{contour_deformation}) and (\ref{lambda_su}), paying attention to the phases appearing in the manipulations. First consider 
\bea
I=\int d^4x~e^{-ix^2/4s} \equiv J^3 K, ~~~~s>0,
\label{main_int}
\eea
where 
\bea
J=2\int\limits_0^\infty dx_1~e^{ix_1^2/4s},~~~~K=2\int\limits_0^\infty dx_0~e^{-ix_0^2/4s}.
\eea
For $J$ we put $x_1^2/4s =u$ to get
\be
J = 2\sqrt{s}\int\limits_0^\infty du~u^{-1/2}~e^{iu}.\nn
\ee
To avoid oscillations in the integrand, we take the contour of Fig.\ref{contour_mf1} in the first quadrant of the complex $u$ plane, so that the contribution from the quarter circle vanishes. As there is no singularity within and on the contour, the Cauchy formula gives
\be
J = -2\sqrt{s}\int\limits^0_{i\infty} du~u^{-1/2}~e^{iu}.\nn
\ee
If we now put $u=\exp\left(i\pi/2\right)t$ on the integration line along the imaginary axis, we get
\bea
J = 2\sqrt{s}~e^{i\pi/4}\int\limits_0^\infty dt~t^{-1/2}~e^{-t} = e^{i\pi/4}~\sqrt{4\pi s},
\label{J_int}
\eea
the integral being the familiar Gamma function $\Gamma(1/2) = \sqrt{\pi}$. The integral $K$ can be evaluated in the same way, taking the contour of Fig.\ref{contour_mf2} in the fourth quadrant, making the contribution of the quarter circle to vanish. We then get 
\bea
K=e^{-i\pi/4}~\sqrt{4\pi s}.
\label{K_int}
\eea
Putting the results (\ref{J_int}) and (\ref{K_int}) in (\ref{main_int}) we get
\be
I=i(4\pi s)^2.
\ee

Next we consider the Fourier transform Eq.(\ref{lambda_su}) 
\bea
\Lambda(s,u,q_\shortparallel,q_\perp) &=& e^{-im^2s}\int d^4 x ~e^{i q\cdot x}\left[-\alpha x_\perp^2 +\beta x_\shortparallel^2+16 i (\gamma +\delta)\right]\exp\left(i\gamma x_\perp^2 - i\delta x_\shortparallel^2\right).\nn
\eea
Here the basic integrals are 
\bea
L_1 &=& \int\limits_{-\infty}^{+\infty} dx_\perp^2~e^{-iq_\perp\cdot x_\perp + i\gamma x_\perp^2},\nn\\
L_2 &=& \int\limits_{-\infty}^{+\infty} dx_0~e^{iq_0 x_0 - i\delta x_0^2},\nn\\
L_3 &=& \int\limits_{-\infty}^{+\infty} dx_3~e^{-iq_3 x_3 + i\delta x_3^2},\nn
\eea
in terms of which the Fourier transform may be written as
\bea
\Lambda = i\alpha \frac{dL_1}{d\gamma}L_2L_3 + i\beta \left(L_1\frac{dL_2}{d\delta}L_3 + L_1L_2\frac{dL_3}{d\delta}\right) + 16~i~(\gamma+\delta)~L_1L_2L_3.
\label{ft_lambda}
\eea
The basic integrals are generally of the same form as $J$ and $K$, if we complete the squares in the exponents. Thus 
\bea
L_1 &=& e^{-iq_\perp^2/4\gamma} \int\limits_{-\infty}^{+\infty} dx_\perp^2~\exp\left[i\gamma\left(x_\perp-\frac{q_\perp}{2\gamma}\right)^2\right],\nn\\
&=& e^{-iq_\perp^2/4\gamma} \pi\int\limits_{0}^{\infty} dt~e^{i\gamma t},\nn
\eea
on substituting $x_\perp \rightarrow x_\perp + q_\perp/2\gamma$ and using polar coordinates. Taking a contour in the first quadrant, we get
\bea
L_1 = e^{i\pi/2} ~\frac{\pi}{\gamma}~e^{-iq_\perp^2/4\gamma}.
\eea
In the same way we can evaluate the integrals $L_2$ and $L_3$ by taking contours respectively in the fourth and first quadrants,
\bea
L_2 &=& e^{-i\pi/4} ~\frac{\sqrt{\pi}}{\sqrt{\delta}}~e^{iq_0^2/4\delta},\\
L_3 &=& e^{i\pi/4} ~\frac{\sqrt{\pi}}{\sqrt{\delta}}~e^{-iq_3^2/4\delta}.
\eea
Putting these values of integrals $L_i,~i=1,2,3$, in (\ref{ft_lambda}) we get $\Lambda$ as given by Eq.(\ref{lambda_fap}) in the text.

\end{document}